\documentstyle[aps, graphicx, twocolumn]{revtex}
\begin{document}

\title{
Emergence of a dominant unit in a network of chaotic units\\
 with delayed synaptic change}
\author{Junji Ito\footnote{E-mail:ito@complex.c.u-tokyo.ac.jp}}
\address{
Department of\\
Pure and Applied Sciences,\\
College of Arts and Sciences,\\
University of Tokyo\\
3-8-1 Komaba, Meguro,\\
Tokyo 153 Japan\\
}
\author{Toru Ohira\footnote{E-mail:ohira@csl.sony.co.jp}}
\address{
Sony Computer Science Laboratory\\
3-14-13 Higashi-gotanda, Shinagawa,\\
 Tokyo 141, Japan\\
}
\date{\today}
\maketitle
\begin{abstract}
We study here a model of globally coupled units 
with adaptive interaction weights 
which has a delay in the updating rule. 
Simulations show that the model with such delayed synaptic change
exhibits dynamical self-organization of network structure.
With suitably chosen parameters,
``dominant'' unit emerges spontaneously,
in the sense that the connections from such unit 
to almost all of the other units are especially strengthened.
Such weight structure facilitates the coherent activity among units.
\end{abstract}

\vspace{0.5cm}

Self-organization associated with neural networks 
has been one of the main research focus [1]. 
It has been studied from both dynamical aspects such as firing patterns,
and structural aspects like strength of synaptic connections.
The main principle for many self-organizing characteristics
is the Hebbian rule. 
Ito and Kaneko [2] has recently proposed 
a model of globally coupled chaotic units 
with the Hebbian type synaptic updating rule.
The striking finding from the simulation of the model is
that with an external input unit,
such network model self-organizes into the layered structures.
In this model, some layers can be defined
according to the distance from the external input unit
measured along the connection between units.
The notable point is that such structure emerges spontaneously
without any predetermined instruction,
except which unit the external input is given to.
The model has given a possible starting point 
for the mechanism of self-organization of the layered structure.

The main theme of this Letter is the study of this model 
when we incorporate a delay in the updating rule of synaptic weight.
The change of weight lags 
behind the change of the state of the model neuron unit 
reflecting a delay between presynaptic and postsynaptic process [1].
In biological information processing, 
delay is an important factor, 
and the effect of delay on dynamical systems is
perceived as a source of more complexity [3,4,5,6]. 
Our numerical solution indeed shows that 
delay contributes to phase of more mixed dynamical patterns in the network.
When the model structure is concerned, however, 
the delay induces a peculiar dynamical order. 
Without delay, we have designated a single input neuron 
which is supplied with an external fixed input.
Some structural order is induced with that neuron as a ``core''.
With delay, such a ``core'' or ``dominant'' neuron emerges spontaneously. 
In other words, some structural coherence is observed 
due to delay without an external fixed input.
Ohira and Sato [7] has recently proposed a simple model to
induce a regular spiking patterns using delay.
There are also studies 
that delay could lead to suppression of complex dynamics [8,9]. 
Our model here can be considered as one of such examples 
exhibiting an effect of delay toward order.
From the neural network point of view, 
this model could lead to a possibility of emergence 
of neural network structure without any driving external input.

As an abstract model for neural information processing, 
we adopt a simple dynamical system model 
that has the following two properties:
(1) global connections between nonlinear units 
that can exhibit chaotic and other dynamic behavior, 
and 
(2) plastic change of the connections between units
that depend not only on the states of the two units to which they correspond
but also globally on all the other units.

Specifically we employ a globally coupled map (GCM) [10] 
with plastic couplings. 
Although, in a conventional GCM, couplings between units are fixed,
we need to introduce a process 
corresponding to the change of the synaptic weight.
Here we introduce variable connection strengths between units.
In some applications of GCM to neural networks,
a unit of the GCM corresponds to a single neuron [11,12].
We do not adopt such a representation in our model.
Instead we study neural dynamics at a coarse-grained level,
by regarding the dynamics of each unit of the GCM 
as a collective variable representing an ensemble of neurons.
Here, we are interested in the universal behavior of systems 
which possess the features described above, 
rather than the phenomena specific 
to any particular choice of the unit dynamics.  

Each unit in the system is taken to be the logistic map.
The output $x_n$ at the time step $n$ is given as follows:
\[
x_{n+1} = a \cdot x_{n} ( 1 - x_{n} ).
\]
$a$ is the parameter representing the nonlinearity of the map,
which can take the value between 0 and 4.

With units of this type,
we consider the following network model:
\[
x^{i}_{n+1} = a \cdot y^{i}_{n} ( 1 - y^{i}_{n} )
\]
\[
y^{i}_{n} = ( 1 - c ) x^{i}_{n} + c \sum_{j=1}^{N} \varepsilon^{ij}_{n}
x^{j}_{n}.
\]
Here $x^{i}_n$ and $y^{i}_n$
are the output and the state variable of the unit $i$,
respectively, at the time step $n$.
The variable $\varepsilon^{ij}_n$ is time-dependent synaptic weight
of the coupling from unit $j$ to unit $i$.
$N$ is the number of the units,
and $c$ is the parameter which represents
the strength of the influence of other units.

In general, for systems of the type we study, 
the variation of the connections has been described by Hebbian-type dynamics.
It is not straightforward 
to introduce the coupling dynamics in the present case, 
since the state $x^i_n$ refers not to a neuron but to a collective state.
We extract the essence of the above dynamics of synapses,  
and include it in our abstract GCM model.
Here we consider the dynamics for the synaptic strength described by

\[
\varepsilon^{ij}_{n} = \frac{\tilde{\varepsilon}^{ij}_{n}}{ \sum_{j=1}^{N}
\tilde{\varepsilon}^{ij}_{n}},
\]
\[
\tilde{\varepsilon}^{ij}_{n+1} = \cases{ [1+\cos \pi (x^{j}_{n -
\tau } - x^{i}_{n})] \varepsilon^{ij}_{n} & (for $i \ne j$) \cr 0 & (for
$i=j$) \cr }.
\]
$\tau$ is the delay time.
When $\tau = 0$, this updating rule is the same as in the original model
[2].
The `normalization' over all units 
gives a simple representation of the global competition 
for the coupling change.  

As an initial condition, 
all the coupling strengths $\varepsilon^{ij}_0$ are set to be identical,
and the state variables $x^{i}_0$ are drawn from a random number 
between 0 and 1.

In simulations shown below,
we use parameter values
$a = 3.7$, $c = 0.25$, and $\tau = 1$.
These parameter values correspond to partially ordered phase [10]. 
In this phase, 
the dynamics of state variables exhibits chaotic itineracy, 
which is characterized 
by the dynamic change of the effective degrees of freedom.
In a conventional GCM,
this phase is observed for relatively narrow ranges of parameter values.
Here, by the introduction of delay,
partially ordered phase appears in much wider regime in the parameter space.
Hence, the effect of delay produces a richer dynamical behavior.

\begin{figure}[t]
\includegraphics[width=8cm]{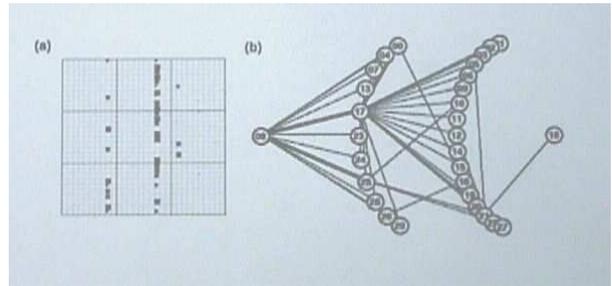}
\caption{
(a) Snapshot of the connection matrix.
The size of the square plotted in i-th row and j-th column
is proportional to the value of $\varepsilon_{ij}$,
and $\varepsilon_{ij}=1$
when the square size is equal to the grid size.
This snapshot is taken from the same session as used in fig.3
at 1200th step.
(b) One example of the graph representation
of the network structure.
Numbers written in circles represent unit indices,
while lines between circles correspond to connection
between units.
This graph is drawn according to the same connection matrix
as shown in (a), with unit 8 as a starting unit,
while another choice of a starting unit alters the graph structure.
Detailed method to draw this graph is written in ref [2].
}
\end{figure}

\begin{figure}[t]
\includegraphics[width=8cm]{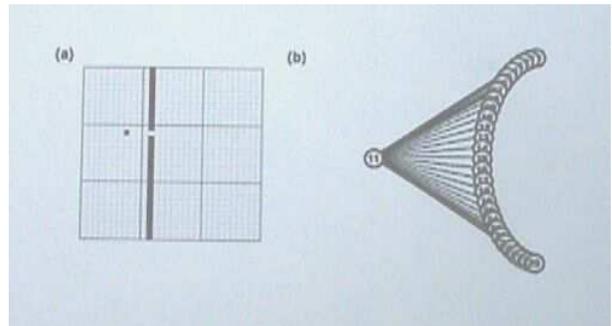}
\caption{
(a) Snapshot of the connection matrix,
taken from the same session as used in fig.3 at 20000th step.
(b) One example of the graph representation
of the network structure, 
drawn according to the same connection matrix
as shown in (a), with unit 11 as a starting unit.
}
\end{figure}

Although the dynamics of state variables exhibit such a complex movement,
the situation is different as we turn our attention 
to the dynamics of the synaptic weights.
Snapshots of the strength of the connection matrices 
$\varepsilon^{ij}_{n}$ at different time step are shown 
in figure 1 and figure 2.
Fig. 1(a) is a snapshot taken at the 1200th time step.
One can easily see that
the stronger connections are concentrated in a few columns.
In this figure, the filled squares in the $i$th column
represents the connection from $i$th unit to the other units,
so fig. 1(a) represents the situation that
the connections from a few units
to almost all the other units are selectively strengthened.
These few units emerged as ``core'' or ``dominant'' units
of the structure of the network.
In figure 1(b), 
one example of the graph representation of the network structure, 
drawn according to the method used in [2], is shown.
The dominance of unit 8 and 17 over the rest of units can be seen.

After sufficient time steps are elapsed,
the connection matrix appears like the state shown in figure 2(a).
In this figure, taken at 20000th time step,
almost all the connections are sited in the only one column,
reflecting an emergence of a single dominant unit.
One connection is sited in the other column
because of the characteristics of our model that
the self connection is meant to be 0.
In fig. 2(b), the graph representation of
the network structure is shown again,
illustrating the prominent dominance of unit 11.
During this stage, the concentration of the coupling strength
to one column is relatively stable
so that such state can sustain for thousands of time steps.

Considering the apparent tendency that
the connection strength would gather to a few columns,
we calculated the summation of the connection strength
over each column at each time step, {\it i.e.},
$\sum_{i=1}^{N}  \varepsilon^{ij}_{n}$,
which represents, in a sense,
the strength of the influence of the unit $i$ on the other units,
and plotted the time series of this value averaged for every 100 steps.
Figure 3 is the plot.
The frequent changes of the dominant core unit
during the earlier stage, and the stable lasting of the dominance
by a particular core unit for up to about 20000 steps 
during the later stage can be seen.

Now let us consider the influence of the appearance
of such dominant unit on the dynamics of state variables.
As described above, in the parameter regime we now consider,
system exhibits the chaotic itineracy which accompanies
the dynamic temporal change of the effective degrees of freedom.
One method to evaluate the effective degrees of freedom is
to calculate the number of clusters with low resolution [10].
Figure 4 is the calculated number of clusters
with three different resolutions.
Prominent declines of the effective degrees of freedom are
observed in two periods, namely,
from the 5000th step to the 8000th step,
and after the 18000th step.
Note that these periods corresponds to the appearance of
the dominant unit, as is shown in fig. 3.

The decline of the effective degrees of freedom implies
the coherence of dynamics of state variables.
To investigate whether such coherent dynamics really happens,
we computed the distribution of $x^{i}_{n}$ 
around the mean value, $\sum_{i=1}^{N} x^{i}_{n} / N$, at each step.
The distribution is calculated for three different periods,
say, 0th-5000th step, 5000th-10000th step, 
and 20000th-25000th step.
Note that the first period corresponds
to the frequent change of the dominant unit,
and the latter two periods to the stable lasting of the dominance.
The results are shown in figure 6(a), 6(b), and 6(c), respectively.
All of these three figure show the peak at the center.
However, the peak is much keener
and the width of the distribution is thinner for the last two figures,
which imply that the coherent activity among units do emerge.

\begin{figure}[t]
\includegraphics[width=10cm]{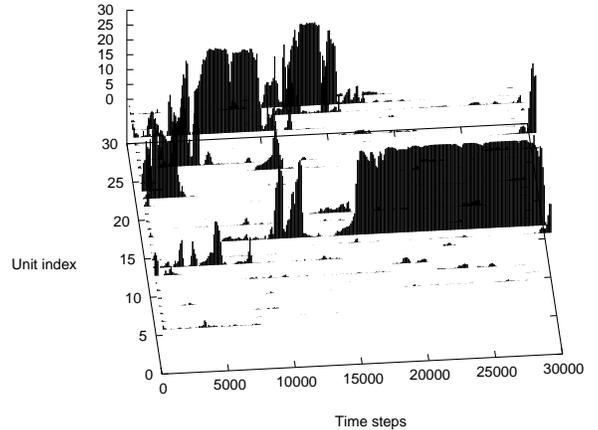}
\caption{
An example of time series of $\sum_{i=1}^{N}
\varepsilon^{ij}_{n}$,
whose values are represented by the lengths of vertical lines.
}
\end{figure}

\begin{figure}[t]
\includegraphics[width=8cm]{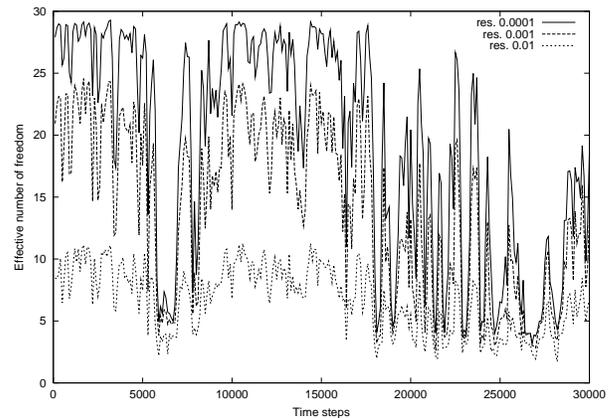}
\caption{
The effective degrees of freedom calculated with
three different resolutions, {\it i.e.}, 0.0001, 0,001, and 0.01.
This plot is obtained using the data of the same session as used in fig.3.
}
\end{figure}

\begin{figure}
\begin{center}
\includegraphics[width=4cm]{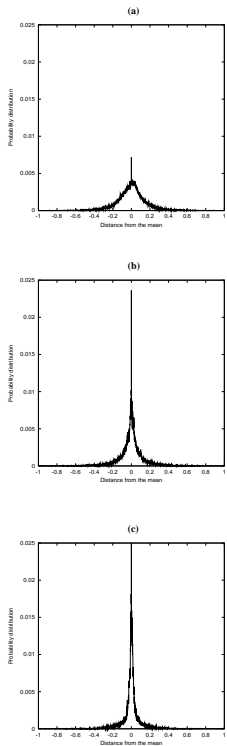}
\end{center}
\caption{
Distribution of $x^{i}_{n}$ around the mean value.
(a) Between 0th and 5000th step.
(b) Between 5000th and 10000th step.
(c) Between 20000th and 25000th step.
}
\end{figure}

There are a couple of points for discussion.
The first point is the dynamical emergence of dominant units 
in the weight structure of the network. 
Without delay, as shown in the previous study [2], 
a core unit was needed to be manually placed 
by representing an external constant input. 
This input unit then became a core unit of structured weight connections. 
Here, as we have shown, 
delay induced the dynamical emergence of such core units.
It is yet to be investigated
whether this model, with such dynamically emerged core units,
can also show a layered structure as in the original model.

Also, we note that there is a dynamical interplay 
between the coherence of weight structure 
and the coherence of activities of units.
As mentioned above,
delay in the synaptic change gives rise to the complex behavior
in the dynamics of state variables.
This dynamics is characterized by temporal change of
the effective degrees of freedom.
The low degrees of freedom implies that
values of state variables are somewhat gathering 
and the high degrees of freedom corresponds to spread values.
Now, let us suppose 
that the system with the relatively low degrees of freedom 
suddenly get the high degrees of freedom.
Since, in this situation,
only a few units have the values near the previous ones,
selective strengthening of connections 
from such a few units to the other units occurs.
This is the mechanism that the dominant unit appears.
The dynamic change of the effective degrees of freedom triggers
the emergence of the dominant units.
Once the dominance of one unit gets sufficiently large,
this unit starts to attract more and more units around it,
since, in this state, 
almost all of the units obey the quite similar rule,
and the dynamics of every unit has to be similar to each other.
This results in almost complete dominance by the only one unit.
However, this dominant state cannot last too long.
Indeed, our numerical simulation shows sudden substitution of a core unit.
The study on the stability of dominant state is one of the further works.

Our model may have rather peculiar points
such as a choice of logistic map units,
and more study is needed to clarify the characteristics of such delay
induced dynamically emerged interplay of coherence between structure and
activities.
We expect, however, that such effect may be of importance with biological
neural networks.

\end{document}